%% file: main-full.tex
\documentclass{article} % For LaTeX2e
\usepackage{nips15submit_e,times}%%%%%%%%%%%%%%%%%% 

\usepackage{hyperref}

\usepackage{amsmath,amssymb,amsthm,amsfonts,latexsym}        
\usepackage{url}
\usepackage{graphicx,subfigure}
\usepackage{bm}
\usepackage{paralist}
\usepackage{color}
\usepackage[normalem]{ulem}
\usepackage[ruled,lined,linesnumbered,noend]{algorithm2e}
\usepackage{appendix}
\usepackage{titlesec}
\titlespacing*{\section}{0pt}{0.8\baselineskip}{\baselineskip}

\makeatletter
\newtheorem*{rep@theorem}{\rep@title}
\newcommand{\newreptheorem}[2]{%
\newenvironment{rep#1}[1]{%
 \def\rep@title{#2 \ref{##1}}%
 \begin{rep@theorem}}%
 {\end{rep@theorem}}}
\makeatother

\makeatletter
\renewcommand{\SetKwInOut}[2]{%
  \sbox\algocf@inoutbox{\KwSty{#2}\algocf@typo:}%
  \expandafter\ifx\csname InOutSizeDefined\endcsname\relax% if first time used
    \newcommand\InOutSizeDefined{}%
    \sbox\algocf@inoutbox{\KwSty{#2}\algocf@typo\textbf{:}~}\setlength{\inoutindent}{\wd\algocf@inoutbox}%
  \else% else keep the larger dimension
    \ifdim\wd\algocf@inoutbox>\inoutsize%
    \sbox\algocf@inoutbox{\KwSty{#2}\algocf@typo\textbf{:}~}\setlength{\inoutindent}{\wd\algocf@inoutbox}%
    \fi%
  \fi% the dimension of the box is now defined.
  \algocf@newcommand{#1}[1]{%
    \ifthenelse{\boolean{algocf@inoutnumbered}}{\relax}{\everypar={\relax}}%
    {\let\\\algocf@newinout\hangindent=\inoutindent\hangafter=1\KwSty{#2}\algocf@typo\textbf{}~##1\par}%
    \algocf@linesnumbered% reset the numbering of the lines
}}%
\makeatother

\newcommand{\eat}[1]{}

\newcommand{\pr}{\mbox {Pr}} 
\newcommand{\vecmu}{\mbox {$\vec{\mu}$}}  
\newcommand{\vectheta}{\mbox {$\vec{\theta}$}}  
\newcommand{\vecp}{\mbox {$\vec{p}$}}  
\newcommand{\hatmu}{\mbox {$\hat{\mu}$}}  
\newcommand{\algo}{\mbox {{$\cal A$}}}  
  
\newcommand{\failure}{\mbox {$\phi$}}  
\newcommand{\gradient}{\mbox {$\Gamma$}}

\newcommand{\argmax}[1]{\mathop{\hbox{argmax}}_{#1}}

\newtheorem{theorem}{Theorem}
\newreptheorem{theorem}{Theorem}

\title{Influence Maximization with Bandits}
\author{
Sharan Vaswani, Laks V.S. Lakshmanan, Mark Schmidt \\
University of British Columbia \\
\texttt{\{sharanv,laks,schmidtm\}@cs.ubc.ca} \\
}

\nipsfinalcopy
\begin{document}

\maketitle
%\twocolumn[
%%\aistatstitle{Influence Maximization with Bandits}
%%\aistatsauthor{Sharan Vaswani \And Laks V.S. Lakshmanan \And Mark Schmidt}
%%\aistatsaddress{University of British Columbia} 
%]

\input{Abstract}
\vspace*{-2ex} 	
\input{The-Introduction-full}

\input{The-RelatedWork}

\input{The-Theory-full}

\input{Regret}

\input{Final-Experiments}
\input{Conclusion}

\input{Supplementary}
\input{Supplementary2}

\newpage
\small
\bibliographystyle{abbrv}
\bibliography{ref}  

\end{document}

%% file: Abstract.tex
\begin{abstract}
%Viral marketing aims to leverage a social network to spread awareness about a specific product in the market through information propagation via word-of-mouth. The problem of influence maximization (IM) is to identify the `best' set of influential users (seeds) which when given free products or discounts will result in the maximum number of people adopting or becoming aware of the product.
 %Most prior work on influence maximization assumes knowledge of influence probabilities between every connected pair of users in the network, or availability of data from which such knowledge can be gained. Since such information is generally not available or is hard to obtain, we drop this assumption by adopting a combinatorial multi-armed bandit paradigm. %We seek to minimize the regret incurred by choosing suboptimal seed sets over multiple IM attempts. We propsose algorithms and establish bounds on their performance. Finally, 
%In this paradigm, IM is conducted over multiple rounds, in each of which a seed set is selected and feedback from the network is incorporated in selecting seeds in the next round. We consider both edge level and node level feedback and establish bounds on the average regret incurred over a campaign consisting of multiple rounds. 
%We demonstrate the effectiveness of the proposed algorithms via a comprehensive set of experiments over four real datasets. 

% MWS: does the below sound reasonable? I aimed to make it as concise/generic and as 'need-to-know' as possible
We consider the problem of \emph{influence maximization}, the problem of maximizing the number of people that become aware of a product by finding the `best' set of `seed' users to expose the product to. Most prior work on this topic assumes that we know the probability of each user influencing each other user, or we have data that lets us estimate these influences. However, this information is typically not initially available or is difficult to obtain. To avoid this assumption, we adopt a combinatorial multi-armed bandit paradigm that estimates the influence probabilities as we sequentially try different seed sets. We establish bounds on the performance of this procedure under the existing edge-level feedback as well as a novel and more realistic node-level feedback. Beyond our theoretical results, we describe a practical implementation and experimentally demonstrate its efficiency and effectiveness on four real datasets.  

% over the rounds of a CMAB game. %The few papers that propose algorithms for learning these probabilities assume the availability of a batch of diffusion cascades and learn the probabilities offline. %(i) learning influence probabilities and (ii) %network exploration, i.e., minimizing the error in learned influence probabilities, and 

\eat{ 
In this paper, we explore the problem of influence maximization in social networks when the graph structure (specifically the influence probabilities) in the network is not fully known. We formulate this problem as a combinatorial multiarmed bandit problem with the arms representing the edges in the network. We propose two related objectives for this problem: 1)minimizing the regret in the spread of the influence maximization process and 2) minimizing the number of cascades needed to learn the influence probabilities in the network. 
We modify the existing multiarmed bandit algorithms namely Upper Confidence Bound (UCB), $\epsilon$-greedy and $\epsilon$-initial for minimizing regret. We use random and strategic exploration algorithms for learning the influence probabilities in an online fashion. We conduct extensive experiments on several large real world datasets to prove the effectiveness and scalability of our approach.}  
\end{abstract}

%% file: The-Introduction-full.tex
\vspace*{-2ex} 
\section{Introduction}
\label{sec:Introduction}
\vspace*{-2ex} 
Viral marketing aims to leverage a social network to spread awareness about a specific product in the market through information propagation via word-of-mouth. Specifically, the marketer aims to select a fixed number of `influential' users (called \emph{seeds}) to give free products or discounts to. The marketer assumes that these users will influence their neighbours and, by transitivity, other users in the social network. This will result in information propagating through the network as an increasing number of users adopt or become aware of the product. %We use \emph{influence spread} to refer to the number of users in the network who have adopted the product. 
The marketer typically has a budget on the number of free samples or discounts that can be given, so she must strategically choose seeds so that the maximum number of people in the network become aware of the product. The goal is to maximize the \emph{spread} of this influence, and this problem is referred to as \emph{influence maximization} (IM)~\cite{kempe2003maximizing}.

\eat{The IM problem needs to be solved each time the marketer wants to spread awareness or encourage adoptions of a new product.} 
In their seminal paper, Kempe et al. \cite{kempe2003maximizing} studied the IM problem under well-known probabilistic information diffusion models including the independent cascade (IC) and linear threshold (LT) models. While the problem is NP-hard under these models, there have been numerous papers on efficient approximations and heuristic algorithms (see Section~\ref{sec:Related-Work}). But prior work on IM assumes that in addition to the network structure, we either know the pairwise user influence probabilities or that we have past propagation data from which these probabilities can be learned. However, in practice the influence probabilities are often not available or are hard to obtain. To overcome this, the initial series of papers following \cite{kempe2003maximizing} simply {\sl assigned} influence probabilities using some heuristic means. 
\eat{
: e.g., assign a fixed small probability such as $0.01$ to all edges, assign values drawn at random from a fixed set such as $\{0.1, 0.01, 0.001\}$, or assign an equal probability of $\frac{1}{N^{in}(v)}$ to all incoming edges of a node $v$, where $N^{in}(v)$ is the set of in-neighbours of $v$. } 
However, Goyal et al. \cite{goyal2011data} showed empirically that learning the influence probabilities from propagation data is critical to achieving seeds and a spread of high quality. 

In this work, we consider the practical situation where even the propagation data may not be available.% or may be proprietary. 
We adopt a combinatorial multi-armed bandit (CMAB) paradigm and consider an IM campaign consisting of multiple rounds (as in another recent work~\cite{chen2014combinatorial}). Each round amounts to an IM attempt and incurs a \emph{regret} in the influence spread because of the lack of knowledge of the influence probabilities. We seek to minimize the accumulated regret incurred by choosing suboptimal seed sets over multiple rounds. 
%We select seeds based on available knowledge about influence probabilities, beginning with no knowledge (other than the graph structure). As we do more rounds of seed selections, based on observations of rewards from previous rounds, our knowledge improves. 
A new marketer may begin with no knowledge (other than the graph structure) and at each round we can choose seeds that improve our knowledge and/or that lead to a large spread, leading to a class \emph{exploration-exploitation} trade-off. (An alternative to minimizing the regret can be to just learn the influence probabilities in the network as efficiently as possible. This is referred to as pure exploration~\cite{bubeck2011pure,chen2014combinatorialexp} and we briefly explore it in Appendix C.)
%We can either leverage this knowledge to choose better seeds in future rounds or choose seeds that would further our knowledge, leading to a classic \emph{exploration-exploitation} trade-off. 
%Practically, a new marketer who has just entered the market and has no knowledge about the network will try and learn about it through a trial and error procedure (`exploration'). These exploration rounds are therefore like an investment and the knowledge gained in this phase can be used to generate revenue in the subsequent rounds (`exploitation'). 
%In this work, our main contribution is to formulate IM with unknown influence probabilities as a (combinatorial) MAB problem (as in ~\cite{chen2014combinatorial}). 
As in prior work, we first consider ``edge-level'' feedback where we assume we can observe whether influence propagated via each edge in the network (Section~\ref{sec:Theory}). However, we also propose a novel ``node-level'' feedback mechanism which is more realistic (Section~\ref{sec:NL}): it only assumes we can observe whether each node became active (e.g., adopted a product) or not, as opposed to knowing who influenced that user.
%We use two different feedback mechanisms to model the information obtained from previous seed selections (Section~\ref{sec:Theory}). We first formulate our problem using ``edge level'' feedback, which assumes we can observe whether influence propagated via each edge in the network. We also propose a novel ``node level'' feedback mechanism which is more realistic in practice: it only assumes we can observe whether each node became active (e.g., adopted a product) or not, as opposed to who activated that node. 
We establish bounds on the regret achieved by the algorithms under both kinds of feedback mechanisms. We further present regret minimization algorithms (Section \ref{sec:Regret}) and conduct extensive experiments on four real datasets to evaluate the effectiveness of the proposed algorithms (Section~\ref{sec:Experiments}). All proofs appear in the supplementary part, which also explores the effect of prior on performance and discusses the alternative objective of network exploration.

%% file: The-RelatedWork.tex
\section{Motivation and Related Work}
\label{sec:Related-Work}
%IM in general 
\vspace*{-2ex} 
\noindent 
{\bf Influence Maximization}: We model a social network as a probabilistic directed graph $G = (V,E,P)$ with nodes $V$ representing users, edges $E$ representing connections/contacts, and edge weights $P: E\rightarrow [0,1]$. The influence probability $P(u,v)$ represents the probability with which user $v$ will perform an action given that it is performed by $u$. A stochastic diffusion model $D$  governs how information spreads from nodes to their neighbours in the network. Given a seed set $S$, the expected number of nodes of $G$ influenced by $S$ under the model $D$, denoted $\sigma_D(S)$ (just $\sigma(S)$ when $D$ is obvious from context), is called the (expected) \emph{influence spread} of $S$. Given $G$ and a budget $k$ on the number of seeds to be selected, IM aims to find the seed set $S$ of size $k$ which will lead to the maximum influence spread $\sigma(S)$ under $D$,
\begin{equation}
S^* = \argmax{|S| \leq k} \sigma(S).
\label{eq:IM}
\end{equation}
Although IM is NP-hard under standard diffusion models, the expected spread function $\sigma_D(S)$ is monotone and submodular. Solving Eq. (\eqref{eq:IM})  thus reduces to maximizing a submodular function under a cardinality constraint, a problem that can be solved to within a $(1-1/e)$-approximation using a greedy algorithm~\cite{nemhauser1978analysis}. There have been a variety of extensions including development of scalable heuristics, alternative diffusion models, and scalable approximation algorithms~\cite{chen2009efficient}, \cite{wang2012scalable}, \cite{leskovec2007cost},  \cite{goyal2011simpath}, \cite{goyal2011data}, \cite{Tang2014Influence}, \cite{Tang2015TIM+}). We refer the reader to \cite{chen2013information} for a detailed survey. Most work on IM assumes knowledge of the influence probabilities, but there is a growing body of work on learning the influence probabilities from data. Typically, the data is a set of  diffusions (also called \emph{cascades}) that happened in the past, specified  in the form of a log of actions by network users. Learning influence probabilities from available cascades has been used discrete-time models~\cite{saito2008prediction,goyal2010learning,netrapalli2012learning} and continuous-time models~\cite{gomez2010inferring}. However, in many real datasets the cascades are not available. For these datasets, we can't even use these existing approaches for learning the influence probabilities.
%We notice that in several real datasets, only the social network data is available and no information about cascades is available. 
% renders the existing approaches for learning influence probabilities inapplicable and motivates us to consider a new framework for IM using only the network structure.  

{\bf Multi-armed Bandit}: The stochastic multi-armed bandit (MAB) paradigm was first proposed in~\cite{lai1985asymptotically}. In the traditional framework, there are $m$ arms each of which has an unknown reward distribution. The bandit game proceeds in rounds and in every round $s$, an arm is played and a corresponding reward is generated by sampling the reward distribution for that arm. This game continues for a fixed number of rounds $T$. 
%Some of the arms result in higher rewards than others. %Initially, the reward distribution of every arm is unknown to the bandit. 
%There are two competing goals for bandit games: 
Our goal is to minimize the regret resulting from playing suboptimal arms across rounds ({\sl regret minimization}). This results in a trade-off between \emph{exploration} (sampling arms to learn about them) and \emph{exploitation} (pulling the arm which we think gives the highest expected reward). Auer et al. \cite{auer2002finite} proposed algorithms which can achieve the optimal regret of $\mathcal{O}(\log(T))$ over $T$ rounds. The \emph{combinatorial} multiarmed bandit paradigm is an extension where we can pull a set of arms (a `superarm') together~\cite{gai2012combinatorial,chen2013combinatorial,gai2010learning,anantharam1987asymptotically}. The subsequent reward could be a linear \cite{gai2012combinatorial} or non-linear \cite{chen2013combinatorial} combination of the individual rewards. Gai et al. \cite{gai2012combinatorial} and Chen et al. \cite{chen2013combinatorial} consider a CMAB framework with access to an approximation oracle to find the best (super)arm to be played in each round. Gopalan et al. \cite{gopalan2014thompson} propose a Thompson sampling based algorithm for regret minimization. Chen et al.~\cite{chen2014combinatorial} introduce the notion that triggering superarms can also probabilistically trigger other arms. They target both ad placement on web pages and viral marketing applications under semi-bandit feedback~\cite{audibert2011minimax}. They propose an algorithm based on the upper confidence bound (UCB) called combinatorial UCB (CUCB) for obtaining an optimal regret of $\mathcal{O}(\log(T))$. However, they assume the often-unrealistic ``edge-level'' feedback and did not experimentally test their algorithm. In contrast, in this work we consider  more realistic ``node-level" feedback and show that our proposed algorithm gives strong empirical performance. More recently, Lei et al. \cite{lei2015online} studied the related, but very different, problem of maximizing the distinct number of nodes activated across rounds. However, they assume edge-level feedback, do not establish any quality guarantees, and do not theoretically compare performance of their approach with that achievable when influence probabilities are known. Further, they synthetically assigned ``true" probabilities while we also test our algorithms on datasets where these probabilities are learned from real data.

%% file: The-Theory-full.tex
\vspace*{-2ex} 
\section{CMAB Framework for IM}
\label{sec:Theory}
\vspace*{-2ex} 
\subsection{Review of CMAB} 
\label{sec:cmab-rev} 
\vspace*{-2ex} 
The CMAB framework consists of $m$ base arms. Each arm $i$ is associated with a random variable $X_{i,s}$ which denotes the outcome or reward  of triggering arm $i$ arm on trial $s$. The reward $X_{i,s}$ is bounded on the support $[0,1]$ and is independently and identically distributed according to some unknown distribution with mean $\mu_{i}$. In each of the $T$ rounds, a superarm $A$ (a set of base arms) is played, which triggers all arms in $A$. In addition, some of the other arms may get probabilistically triggered. Let $p^{i}_A$ denote the triggering probability of arm $i$ if the superarm $A$ is played (observe that $p_A^i = 1$ for $i \in A$). The reward obtained in each round $s$ can be a (possibly non-linear) function of the rewards $X_{i,s}$ for each arm $i$ that gets triggered in that round. Let $T_{i,s}$ denote the total number of times an arm $i$ has been triggered at round $s$. For the special case of $s=T$, we use the notation $T_i := T_{i,T}$. Each time an arm $i$ is triggered, we use the observed reward to update its mean estimate $\hatmu_i$. The superarm that is expected to give the highest reward is selected in each round by an oracle $O$. The oracle takes as input the current mean estimates $\vec{\hatmu} = (\hatmu_{1}, ..., \hatmu_{m})$, and outputs an appropriate superarm $A$. In order to accommodate intractable problems, the framework of \cite{chen2013combinatorial, chen2014combinatorial} assumes that the oracle provides an $(\alpha, \beta)$-approximation to the optimal solution; the oracle outputs with probability $\beta$ a superarm $A$ such that it attains an $\alpha$ approximation to the optimal solution.%, based on the current mean estimate $\vec{\hatmu}$. 

\vspace*{-2ex} 
\subsection{Adaptation to IM} 
\label{sec:cmab-im} 
\vspace*{-2ex} 
Though our framework is valid for any discrete time diffusion model, we will assume the IC diffusion model in our discussion. This model uses discrete steps. At time $t=0$, only the seed nodes are active. Each active node $u$ gets one attempt to influence/activate each of its inactive out-neighbours $v$ in the next time step. This activation attempt succeeds with influence probability $p_{u,v} := P(u,v)$. An edge along which an activation attempt succeeded  is said to be \emph{live}, and other edges are said to be \emph{dead}. 
%We refer to the in-neighbors of a node as its parents. 
At a given time $t$, an inactive node $v$ may have multiple parents which activated at time ($t-1$). This set of parents are capable of activating $v$ at time $t$ and we refer to them as the active parents of $v$ at time $t$. There can be $2^{\vert E \vert}$ (each edge can be live or dead) possible samples (referred to as possible worlds in the IM literature) of the probabilistic network $G$. The sample corresponding to the diffusion in the real world is referred to as the ``true'' possible world and results in a labelling of nodes as influenced (active) or not influenced. The actual spread is the number of nodes reachable from the selected seed nodes in the true possible world and we denote it by $\bar{\sigma}$. 
\begin{table*}
\begin{tabular}{| l | c | c | }
\hline
CMAB & Symbol & Mapping to IM \\
\hline
Base arm & $i$ & Edge ($u,v$) \\
\hline
Reward for arm $i$ in round $s$ & $X_{i,s}$ & Status (live / dead) for edge ($u,v$) \\ 
\hline
Mean of distribution for arm $i$ & $\mu_{i}$ & Influence probability $p_($u,v$)$ \\
\hline
Superarm & $A$ & Union of outgoing edges $E_{S}$ from nodes in seed set $S$ \\
\hline
No. of times $i$ is triggered in $s$ rounds & $T_{i,s}$ & No. of times $u$ becomes active in $s$ diffusions \\
\hline
Reward in round $s$ & $r_{s}$ & Spread $\bar{\sigma}$ in the $s^{th}$ IM attempt \\
\hline
\end{tabular}
\caption{Mapping of the CMAB framework to IM}
\label{tab:mapping}
\end{table*}
Table~\ref{tab:mapping} gives our mapping of the various components of the CMAB framework to IM. Note that since each edge can be either live or dead in the true diffusion, $X_{i,s} \in \{0,1\}$ and we can assume a Bernoulli distribution on these values. We describe the CMAB framework for IM in Algorithm~\ref{algo:CMAB-for-IM}. In each round, the regret minimization algorithm $\algo$ selects a seed set $S$ with $\vert S \vert = k$ and plays the corresponding superarm $E_{S}$. $S$ can be selected either randomly (EXPLORE) or by solving the IM problem with the current influence probability estimates $\vec{\hatmu}$ (EXPLOIT). The details for solving Eq. (\ref{eq:IM}) are encapsulated in the oracle $O$ which takes as input the graph $G$ and $\vec{\hatmu}$, and outputs a seed set $S$ under the cardinality constraint $k$. For the case of IM, $O$ constitutes a $(1 - \frac{1}{e},1 - \frac{1}{|E|})$-approximation oracle~\cite{chen2014combinatorial}. Notice that the well-known greedy algorithm used for IM can serve as such an oracle. Once the superarm is played, information diffuses in the network and a subset of network edges become live which leads to a subset of nodes becoming active. The reward $X_{i,s}$ for these edges is $1$. Note that the reward $\bar{\sigma}(S)$ is the number of active nodes at the end of the diffusion process and is thus a non-linear function of the rewards of the triggered arms (edges). After observing a diffusion, the mean estimate $\vec{\hatmu}$ vector needs to updated. In this context, the notion of a \emph{feedback mechanism} $M$ plays an important role. It characterizes the information available after a superarm is played. This information is used to update the model to improve the mean estimates (UPDATE in Algorithm~\ref{algo:CMAB-for-IM}). 
Let $S^{*}$ be the solution to Eq. (\ref{eq:IM}) and let $\sigma^* = {\sigma}(S^{*})$, the optimal expected spread. Since IM is NP-hard, even if the true influence probabilities are known, we can only hope to achieve an expected spread of $\alpha\beta \sigma^*$, where $\alpha = 1 - \frac{1}{e}$ and $\beta = 1 - \frac{1}{|E|}$. We let $S_s$ be the seed set chosen by $\algo$ in round $s$. The regret incurred by $\algo$ is then defined by
\vspace{-2ex}
\begin{equation}
Reg(\mu,\alpha,\beta) = T\alpha\beta \sigma^* - \mathbb{E}_S \big[ \sum_{s = 1}^{T} \bar{\sigma}(S_{s}) \big]
\label{eq:IM-regret}
\end{equation}
\vspace{-0.5ex}
where the expectation is over the randomness in the seed sets output by the oracle. 

The usual feedback mechanism is the \emph{edge-level feedback} proposed by~\cite{chen2013combinatorial}, where we assume that we know the status (live or dead?) of each triggered edge in the ``true'' possible world. The mean estimates of the arms distributions can then be updated using Eq. (\ref{eq:EL})
\vspace{-2ex}
\begin{equation}
\hat{\mu_{i}} = \frac{\sum_{s = 1}^{t}X_{i,s}}{T_{i,t}}
\label{eq:EL}
\end{equation} 
\vspace*{-2ex}
\begin{algorithm}[t!]
\SetKwInOut{Comment}{}
\SetKwInOut{Output}{Output}
\caption{\textsc{CMAB framework for IM}(Graph $G$ = ($V,E$), budget $k$, Feedback mechanism $M$, Algorithm $\algo$)}\label{algo:CMAB-for-IM}
Initialize $\vec{\hatmu}$ \;
$\forall i$ initialize $T_{i} = 0$ \;
\BlankLine
\For{$s = 1 \rightarrow T$ } {
IS-EXPLOIT is a boolean set by algorithm $\algo$  \;
\If { IS-EXPLOIT } 
{ $E_{S}$ = EXPLOIT($G$,$\vec{\hatmu}$,$O$,$k$) } 
\Else { $E_{S}$ = EXPLORE($G$,$k$) } 
Play the superarm $E_{S}$ and observe the diffusion cascade $c$ \;
$\vec{\hatmu}$ = UPDATE($c$,$M$) \;
}
\end{algorithm}
\vspace*{-2ex}
\section{Node-Level Feedback}
\label{sec:NL} 
\vspace*{-2ex}
Edge-level feedback is often not realistic because success/failure of activation attempts is not generally observable. 
%This motivates us to propose the more realistic node level feedback, made precise below.  
Unlike the status of edges, it is quite realistic and intuitive that we can observe the status of each node: did the user buy or adopt the marketed product? 
%We call this \emph{node-level feedback}. Clearly, compared to edge level feedback, node level feedback makes a weaker and more realistic assumption. 
While this is a more realistic assumption, the disadvantage node-level feedback is that updating  the mean estimate for each edge is more challenging.
This is because we do not know which active parent activated the node, or when it was activated. That is, we have a credit assignment problem.
%To see this, consider a node $v$ which becomes active at time $t$. Let $u_1, ..., u_K$ be the active parents of $v$. (see Figure~\ref{fig:th1-syn-network})
%\begin{figure}[ht]
%\centering
%\includegraphics[scale=0.3]{./figs/th1-syn-network.png}
%\caption{Example for Node Level Feedback}
%\label{fig:th1-syn-network}
%\end{figure}
Under edge-level feedback, we assume that we know the status of each edge  $(u_j, v),$ $1\le j\le K$ and use it to update mean estimates. Under node-level feedback, any of the active parents may be responsible for activating a node $v$ and we don't know which, leading to a credit assignment problem. We describe two ways to resolve this problem. 

\vspace*{-2ex}
\subsection{Maximum Likelihood Approach} 
\label{sec:NLN} 
\label{sec:ne-mle}
\vspace*{-2ex} 
An obvious way to infer the edge probabilities given the status of each node in the cascade is to use maximum likelihood estimation (MLE). We use an MLE formulation similar to those proposed in~\cite{netrapalli2012learning,saito2008prediction}. These papers describe an \emph{offline} method for learning influence probabilities, where a fixed set of past diffusion cascades is given as input. A diffusion cascade captures how information spreads in the network and contains information about if and when each node became active in the diffusion. The log-likelihood function for a given set of cascades $C$ is given by: 
\begin{equation}
\log{L(\vecp)} = \sum_{c = 1}^{C} \sum_{v\in V}\log{L^{c}_{v}(\vecp)}
\label{eq:ml-formulation}
\end{equation}
where $L^{c}_{v}(\vecp)$ models the likelihood of observing the cascade $c\in C$ w.r.t. node $v$, given the influence probability estimates $\bar{p}$. Let $t_{u}^{c}$ be the timestep in the diffusion process at which node $u$ becomes active in cascade $c$. If $p_{u,v}$ is the influence probability of the edge ($u,v$) and $N_{in}(v)$ is the set of incoming neighbours of $v$, $\log{L^{c}_{v}(\vecp)}$ under the IC model can be written as follows:
\begin{equation}
\log{L^{c}_{v}(\vecp)} = \sum_{u \in A^{c}}\ln{(1 - p_{u,v})} + \ln{\left(1 - \prod_{u \in B^{c}}(1 - p_{u,v})\right)}
\label{eq:likelihood}
\end{equation}
Here, $A^{c} = \{u \in N_{in}(v):t_{u}^{c} \leq t_{v}^{c} - 2\}$ and $B^{c} = \{u \in N_{in}(v):t_{u}^{c} = t_{v}^{c} - 1\}$. The first term corresponds to unsuccessful attempts by active parents to activate node $v$, whereas the second term corresponds to the successful activation attempts. Using the transformation 
$\theta_{u,v} = - \ln(1 - p_{u,v})$, Eq. (\ref{eq:likelihood}) becomes 
\begin{equation}
\log{L^{c}_{v}(\vectheta)} = - \sum_{u \in A^{c}} \theta_{u,v} + \ln{\left(1 - \exp{\left(- \sum_{u \in B^{c}} \right) \theta_{u,v}} \right)}
\label{eq:modified-likelihood}
\end{equation}
It can be verified that the log-likelihood function given in Eq. (\ref{eq:modified-likelihood}) is convex. It is also separable across nodes and can be minimized independently for each node using methods like gradient descent. In our setting, we don't have a batch of available cascades but generate cascades on the fly. We can store the generated cascades and use these to find the maximum likelihood estimator for each node in every round of an IM campaign in our bandit framework. As a consequence of observing just node statuses, we incur error in the inferred rewards for each arm, which we characterize next. All proofs appear in Appendix A. 
\begin{theorem}
Let $p^{E}_{u_{i},v}$ and $p^{N}_{u_{i},v}$ resp. denote the probability estimates learned from edge-level and node-level feedback using maximum likelihood. We have: 
Let $p^{E}_{u_{i},v}$ and $p^{N}_{u_{i},v}$ resp. denote the probability estimates learned from edge-level and node-level feedback using maximum likelihood. We have: 
\begin{flalign*}
|p^{N}_{u_{i},v} - p^{E}_{u_{i},v}| \leq \max \bigg( & 1 - p^{E}_{u_{i},v} - \frac{1}{(p^{E}_{u_{i},v} + \failure)(P_{max})}, p^{E}_{u_{i},v} - 1 + \frac{1}{(p^{E}_{u_{i},v} + \failure)(P_{min})} \bigg) 
\end{flalign*}
where $\failure$ is the fraction of cascades  in which edge $(u_{i},v)$ is dead, over those where $v$ is active, and $P_{max}$ and $P_{min}$ are the upper bound and lower bounds on the quantity $\prod_{j \in B^{c} j \ne i} (1 - p_{u_{j},v}^{N})$. 
\label{th:MLE-error}
\end{theorem}
This result bounds the price we pay in terms of error, for adopting a the more realistic node-level feedback over edge-level feedback mechanism. 

%\note[Laks]{We should add a line interpreting what this bound means. E.g., is $P_{max}$ the maximum probability that all other active parents fail?} %\note[Laks]{I have cleaned up the notation and made the theorem human readable. $C$ was used for cascades as well as some constant! It's changed to $\failure$. Also reflected in the appendix. You should double check! Also, do we know that $p^{N}_{u_{i},v} \ge p^{E}_{u_{i},v}$ always holds? I put it in $\vert.\vert$ just in case.}  

\subsection{Online optimization}
Unfortunately, the time complexity of the above approach is $\mathcal{O}(\vert E \vert T^{2})$, which doesn't scale to networks with a large number of edges. To mitigate this, we adapt a result from online convex optimization~\cite{zinkevich2003online} for learning the edge probabilities. Zinkevich et.al~\cite{zinkevich2003online} developed an online convex optimization framework for minimizing a sequence of convex functions over a convex set. In our case, we solve an online convex optimization problem for each node in the network. We first describe some notation used in Zinkevich's framework. Given a fixed convex set $F$, a series of convex functions $c_s: F \rightarrow R$ stream in, with $c_s$ being revealed at time $s$. At each timestep $s$, the online algorithm must choose a point $x_s \in F$ {\sl before} seeing the convex function $c_s$. The objective is to minimize the total cost across $T$ timesteps, i.e., ${\it Cost\/}_{on}(T) = \sum_{s\in[T]} c_s(x_s)$, for a given finite $T$. For the offline setting, the cost functions $c_s$ for $s\in[1,T]$ are revealed in advance, and we are required to choose a {\sl single} point $x\in F$ that minimizes ${\it Cost\/}_{off}(T) =\sum_{s\in [T]} c_s(x)$. The {\em loss}\footnote{We use the term loss instead of regret to avoid confusion with the notion of regret in the CMAB framework.} of the online algorithm compared to the offline algorithm is denoted as ${\it Loss\/}(T) = {\it Cost\/}_{on}(T) - {\it Cost\/}_{off}(T)$. Note the above framework makes no distributional assumption on the streaming convex functions. Zinkevich et al. proposed a gradient descent update %(as in Eq. \ref{eq:OCO-update}) 
for choosing the estimates $x_{s}$: 
\begin{equation}
x_{s+1} = x_{s} - \eta_{s}\nabla(c_{s}(x_s))
\label{eq:OCO-update}
\end{equation}
where $\eta_{s}$ is the step size to be used in round $s$ and $\nabla(c_{s}(x_s))$ is the gradient of the cost function revealed at round $s$. He proved that if we use Eq. (\ref{eq:OCO-update}) and set the step size $\eta_{s}$ according to $1 / \sqrt{s}$, the average loss $\frac{Loss(T)}{T}$ goes down as $\mathcal{O}(\frac{1}{\sqrt{T}})$. 
For our setting, we solve an online convex optimization problem for each node in the network. For each node $v$, $x_s$ corresponds to our $\theta_{s}$ variables and $c_s$ corresponds to the negative log-likelihood function $-\log{L^{s}_{v}(\vectheta)}$. Note that we cannot ensure the cascades are i.i.d., making the usual SGD methods inapplicable. The time complexity of this online procedure is $\mathcal{O}(\vert E \vert T)$. We have the following  theorem which extends a similar result of~\cite{zinkevich2003online}. 
Let $\vec{\theta}_{batch}$ be the set of parameters learned offline with the cascades available in batch, and $\vec{\theta}^s$ be the estimate for the parameters in round $s$ of an IM campaign in the CMAB framework, $d_v$ be the in-degree of $v$ and $\theta_{max} = -\ln{(1 - p_{max})}$. 
\begin{theorem}
If we use Eq. (\ref{eq:OCO-update}) for updating $\vec{\theta}$ with $\eta_{s}$ decreasing as $\frac{1}{\sqrt{s}}$, the following  holds:
\begin{equation}
\sum_{s=1}^{T} (L^{s}_{v}(\vectheta_{batch}) - L^{s}_{v}(\vectheta_{s}) \leq \frac{d_v \theta_{max}^{2} \sqrt{T}}{2} + (\sqrt{T} - \frac{1}{2})\gradient^{2}. 
\label{eq:OCO-specific}
\end{equation}
where $\gradient = max_{s\in[T]} ||\nabla(-L^{s}(\vec{\theta}_{s}))||$ is the maximum L2-norm of the gradient of the negative likelihood function over all rounds. 
\label{th:oco}
\end{theorem}
The average loss $\frac{{\it Loss}(T)}{T}$ can be seen to approach $0$ as $T$ increases. This shows that {\sl with sufficiently many rounds $T$, the parameters learned by the online MLE algorithm are nearly as good as those learned by the offline algorithm}. Since there is a one to one mapping between $\theta$ and $p$ values, as $T$ increases, the parameters $\vecp_{s}$ tend to $\vecp_{batch}$ which in turn approach the ``true'' parameters as the size of the batch, $T$, increases. %Hence, the online approach for maximizing the likelihood results in a consistent estimator of the influence probabilities. 

\subsection{Frequentist Approach}
\label{sec:NLF}
In typical social networks, the influence probabilities are very small. To model this special case, we propose an alternative simple and scalable approach. Low influence probabilities cause the number of active parents i.e. $\vert B^{c} \vert$ to be small. We propose a scheme whereby we choose one of the active neighbours of $v$, say $u_i$, uniformly at random, and assign the credit of activating $v$ to $u_i$. The probability of assigning credit to any one of $K$ active parents is $\frac{1}{K}$. That is, edge $(u_{i},v)$ is given a reward of $1$ whereas edges $(u_j,v)$ corresponding to other active parents $u_j, j\ne i$, are assigned a zero reward. We then follow the same update formula as described  for the edge-level feedback model. 
Owing to the inherent uncertainty in node-level feedback, note that we may make mistakes in credit assignment: we may infer an edge to be live while it is dead in the true world or vice versa. We term the probability of such faulty inference, the \emph{failure probability} $\rho$ under node-level feedback. An important question is whether we can bound this probability. This is important since failures could ultimately affect the achievable regret and the error in the learned probabilities. The following result settles this question. 
\begin{theorem}
Let $p_{min}$ and $p_{max}$ be the minimum and maximum true influence probabilities in the network. Consider a particular cascade $c$ and any active node $v$ with $K_{c}$ active parents. The failure probability $\rho$ for under frequentist node-level feedback for node $v$ is characterized by:  
\begin{equation}
\rho \leq \frac{1}{K_{c}}(1 - p_{min})\bigg(1 - \prod_{k = 1, k \neq i}^{K_{c}}[1 - p_{max}]\bigg) + \bigg(1 - \frac{1}{K_{c}}\bigg)p_{max}. 
\label{eq:failure-probability}
\end{equation}
Suppose $\hat{\mu}_{i}^{E}$ and $\hat{\mu}_{i}^{N}$ are the inferred influence probabilities for the edge corresponding to arm $i$ using edge-level and node-level feedback respectively. Then the relative error in the learned influence probability is given by: 
\begin{equation}
\bigg\vert\frac{ \hat{\mu}_{i}^{N} - \hat{\mu}_{i}^{E}}{\hat{\mu}_{i}^{E}}\bigg\vert = \rho \bigg\vert (\frac{1}{\hat{\mu}_{i}^{E}} - 2) \bigg\vert 
\label{eq:error-in-mean}
\end{equation}
\label{th:failure-prob}
\end{theorem}
\vspace*{-5ex}
From Eq. \ref{eq:error-in-mean}, we observe that as $K_{c}$ increases, the error in the mean estimates increases and it is better to use the maximum likelihood approach for credit distribution. In Section~\ref{sec:Experiments}, we empirically find typical values of $p_{max}$, $p_{min}$, and $K_{c}$ on real datasets and verify that the failure probability is indeed small. We also find that the proposed node-level feedback achieves competitive performance compared to edge-level feedback.

%% file: Regret.tex
\vspace*{-2ex}
\section{Regret Minimization Algorithms}
\label{sec:Regret}
\vspace*{-2ex}
As can be seen from Algorithm~\ref{algo:CMAB-for-IM}, the basic components in the framework are the EXPLORE, EXPLOIT and UPDATE subroutines. EXPLORE outputs a random subset $S$ of size $k$ as the seed set, whereas EXPLOIT consults the oracle $O$ and outputs the seed set that (approximately) maximizes the spread according to current mean estimates $\vec{\hatmu}$. UPDATE examines the latest cascade $c$ and updates the parameters using the desired feedback mechanism $M$. Thus UPDATE may correspond to edge-level feedback Eq. (\ref{eq:EL}) or node-level feedback with frequentist update (\ref{sec:NLF}) or node-level feedback with MLE update (\ref{sec:NLN}). In the remainder of this section, we give four ways to instantiate algorithm $\algo$. They  all invoke the subroutines the EXPLORE, EXPLOIT and UPDATE subroutines. 

\textbf{Upper Confidence Bound:} The Combinatorial Upper Confidence Bound (CUCB) algorithm was proposed in~\cite{chen2013combinatorial} and theoretically shown to achieve logarithmic regret under edge-level feedback. The algorithm maintains an overestimate $\overline{\mu_{i}}$ of the mean estimates $\hat{\mu_{i}}$. More precisely, $\overline{\mu_{i}} = \hat{\mu_{i}} + \sqrt{\frac{3 \ln(t)}{2 T_{i}}}$. Exploitation using $\overline{\mu_{i}}$ values as input leads to implicit exploration and is able to achieve optimal regret~\cite{chen2013combinatorial}. 

\textbf{$\epsilon$-Greedy:} Another algorithm proposed in~\cite{chen2013combinatorial} is the $\epsilon$-Greedy strategy. In each round $s$, this strategy performs exploration with probability $\epsilon_{s}$ and exploitation with probability $1 - \epsilon_{s}$. Chen et al. \cite{chen2013combinatorial} show that that if $\epsilon_{s}$ is annealed as $1/s$, logarithmic regret can be achieved under edge-level feedback.  

The regret proofs for both these algorithms rely on the edge-level mean estimates. We obtain node-level feedback mean estimates in terms of edge-level estimates for both the frequentist (Theorem~\ref{th:failure-prob}) and MLE based (Theorem~\ref{th:MLE-error}) approaches. We can use these these estimates and adapt the proofs to characterize the regret. 

\textbf{Thompson Sampling:} Thompson Sampling requires a prior on the mean estimates. After observing the reward in each round, it updates the posterior of the distribution for each edge. For the subsequent rounds, Thompson Sampling generates samples from the posterior of each edge and performs exploitation by using these samples as input to the oracle $O$.  

\textbf{Pure Exploitation:} This strategy performs exploitation in every round. Since we have no knowledge about the probabilities, it results some implicit exploration.

%% file: Final-Experiments.tex
\vspace*{-2ex}
\section{Experiments}
\label{sec:Experiments}
%%%%%%%%%%%%%%%%%%%%%%%%%% 
\vspace*{-2ex}
\noindent
\textbf{Goals:} Our goal is to evaluate the various algorithms with respect to the regret achieved and the error in the influence probabilities learned compared to the true probabilities. In addition, we also report the running times of key subroutines of the algorithms. 
\noindent \\
\textbf{Datasets:} We use 4 real datasets -- Flixster, NetHEPT, Epinions and Flickr, whose characteristics are summarized in Table~\ref{tab:datasets}. Of these, true probabilities are available for the Flixster dataset, as learned  by the Topic aware IC (TIC) model~\cite{barbieri2013topic}. Since true probabilities are not available for the other datasets, we synthetically assign them according to the weighted cascade model~\cite{kempe2003maximizing}: for an edge $(u,v)$, the influence probability is set to $p_{u,v} = \frac{1}{|N^{in}(v)|}$. It is worth noting that the weighted cascade model is commonly used to evaluate influence maximization algorithms whenever true probabilities and diffusion data are unavailable  ~\cite{kempe2003maximizing,lei2015online,chen2010scalable}. 
\begin{table}[!ht]
\centering
\begin{tabular}{ | l | c | r | r | r | }
\hline
Dataset & $\vert V \vert$ & $\vert E \vert$ & Av.Degree & Max.Degree \\
\hline
NetHEPT & 15K & 31K & 4.12 & 64 \\
\hline
Flixster & 29K & 200K & 7 & 186 \\
\hline
Epinions & 76K & 509K & 13.4 & 3079 \\
\hline
Flickr & 105K & 2.3M & 43.742  &  5425 \\
\hline
\end{tabular} 
\caption{Dataset characteristics}
\label{tab:datasets}
\end{table}
\noindent
\vspace*{-2ex}

\textbf{Experimental Setup:} The probability estimates of our algorithms are initialized to $0$ or set according to some prior information (see Appendix B). A run of consists of playing the CMAB game for $T$ rounds. We simulate the diffusion in the network by sampling a deterministic graph from the probabilistic graph $G$ on each round: for the purpose of our experiments, we assume that the diffusion in the real world happened according to this sample. Given a seed set $S$, the real or ``true" spread achieved in a round is the number of nodes reachable from $S$ in the sample. We define the regret incurred in one round as the difference between the true spread of the seed set obtained using the bandit algorithm, and the true spread of the seed set obtained in given the true probabilities. This eliminates the randomness in the regret because of sampling. For our oracle, as well as for batch-mode seed selection, we use the TIM algorithm~\cite{Tang2014Influence}. This is the state of the art algorithm for IM. We use a maximum of $T =1000$ rounds and use $k=50$ (a standard choice in the IM literature). We verified that we obtain similar results for other reasonable values of $k$. To eliminate any randomness in seed-set selection by the oracle $O$, all our results are obtained by averaging across $3$ runs. 

\noindent 
\textbf{Algorithms Compared:} We consider CUCB, Thompson sampling (TS), $\epsilon$-greedy (EG), pure exploitation (PE).\footnote{Abbreviations used in plots are in parentheses.} For CUCB, if the update results in any $\mu_{i}$ exceeding $p_{max}$, as in previous works we reset it back to $p_{max}$~\cite{chen2013combinatorial}. We set $p_{max} = 0.2$ in our experiments, based on the fact that influence probabilities in practice tend to be small~\cite{goyal2010learning}. For $\epsilon$-Greedy,  we found that $\epsilon_{0} = 5$ works well, and we set the exploration parameter on round $s$ to  $\epsilon_{s} = \epsilon_{0}/s$.  To ensure a fair comparison of Thompson sampling with the other methods, we don't use a prior on the probabilities. We do this by only sampling probability estimates for those edges which have been triggered at least once. Specifically, in each round $s$, $\mu_{i}^{s} \sim Beta(\sum_{j = 1}^{s}X_{i},T_{i,s} - \sum_{j = 1}^{s}X_{i})$. 

\textbf{Baseline algorithms:} We use random selection (RAND) and highest degree selection (HIGH-DEGREE) as baseline methods~\cite{chen2010scalable}. 

\textbf{Feedback Mechanisms:} We consider edge-Level (EL), node-Level Frequentist (NLF), and node-Level maximum likelihood (NL-ML). For NL-ML, although we obtained reasonable performance using Zinkevich's framework, we found it to be sensitive to the particular step-size selected. For all our experiments, we thus report results using the Adagrad regret minimization algorithm~\cite{duchi2011adaptive}, which uses a per-variable step-size that roughly reduced by $1/\sqrt{s}$. We set the initial step-size $\eta_{0}$ to $0.85$. We use the `A-M' to refer to algorithm A with feedback mechanism M. For example, EG-EL means $\epsilon$-greedy with edge-level feedback. Looking at all combinations, we test a total of $12$ combinations of algorithms/feedback, plus two baselines which ignore feedback. We next present our experimental results. 

\noindent 
\textbf{Running times:} In order to characterize the running time for the various algorithms, we present the running time for their key components -- EXPLOIT (P), EXPLORE and UPDATE (U), under all three feedback mechanisms. The time complexity UPDATE under all three feedback mechanisms is $\mathcal{O}(\#(triggered\,edges) T)$. EXPLORE takes $0.003$ seconds for selecting $50$ random seeds for any dataset. 
As the \#seeds $k$ and the  true influence probabilities increase, the number of triggered edges increases and UPDATE takes more time (Table~\ref{tab:runtimes}). 
\begin{table}[!ht]
\centering
\begin{tabular}{ | l | l | l | l | l |}
\hline
Dataset & EXPLOIT & \multicolumn{3}{|c|}{UPDATE} \\
\hline
         &        &    EL    &     NL-F  & NL-ML \\
\hline
NetHEPT  &  0.306 &   0.041  &   0.0104  &  0.003 \\
\hline
Flixster &  1.021 &   0.017  &   0.167   &  0.0396  \\
\hline
Epinions &  1.050 &   0.051  &   1.345   &  0.0893  \\
\hline
Flickr   &  1.876 &   0.551  &    0.7984 &  0.037  \\
\hline
\end{tabular} 
\caption{Subroutine times (in sec/round) for $k$ = 50}
\label{tab:runtimes}
\end{table}

\begin{figure*}[ht]
		\centering
        \subfigure[NetHEPT - EL]
        {
        \includegraphics[scale=0.36]{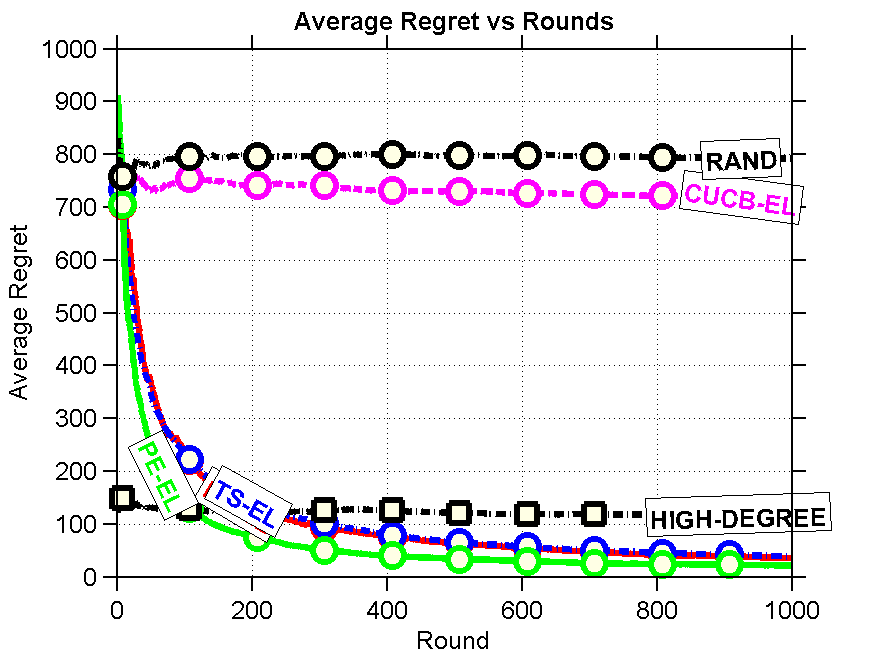}
        \label{fig:nethept-regret-el}
        }
        \subfigure[NetHEPT - NL]
        {
			\includegraphics[scale=0.36]{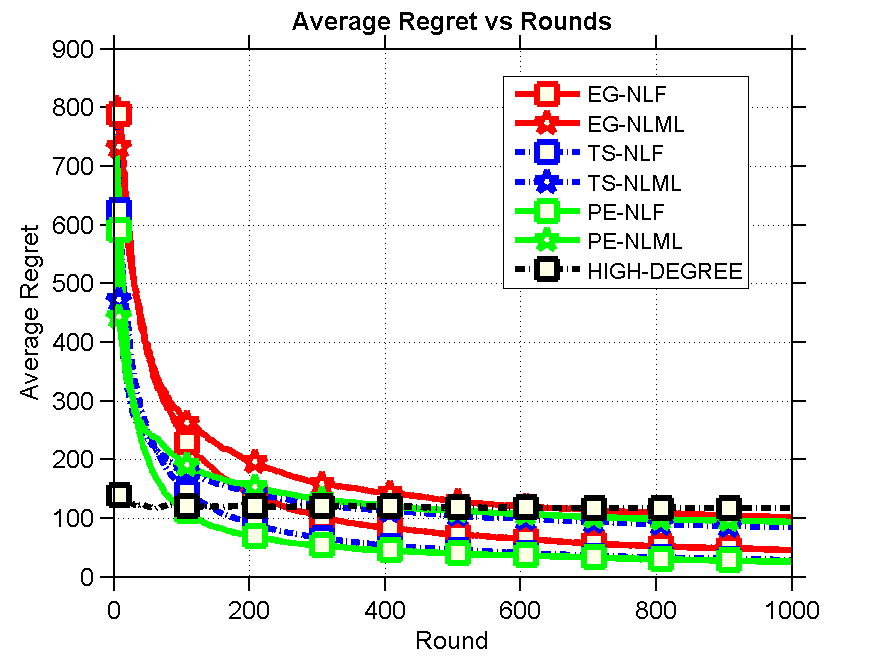}
			\label{fig:nethept-50-regret}
        }
\caption{Regret vs Number of rounds for NetHEPT, $k$ = 50}
\end{figure*}

\noindent
\textbf{Regret Minimization: } We first evaluate the performance of the regret minimization algorithms on NetHEPT assuming edge-level feedback. We plot the average regret, $Regret(T) / T$, as the number of rounds increases. As can be seen from Figure~\ref{fig:nethept-regret-el}, the average regret for PE/EG/TS decreases quickly with the number of rounds. This implies that at the end of $1000$ rounds, the probabilities are estimated by the bandits approaches well enough that they give a comparable spread to using the seeds selected in batch mode given the true probabilities. Pure exploitation achieves the best average regret at the end of $1000$ rounds. This is not uncommon for cases where the rewards are noisy~\cite{chapelle2011empirical}. Initially, with unknown probabilities, rewards are noisy in our case, so exploiting and greedily choosing the best superarm often leads to very good results. Random seed selection has the worst regret, which is constant. For the initial rounds, selecting seeds according to the high degree has a lower regret than other methods. With increasing number of rounds, the influence probabilities become more accurate and the IM oracle $O$ outputs seed sets leading to higher spread than HIGH-DEGREE (this suggest we might reasonably consider a hybrid of these two approaches). We also observe that the regret for CUCB decreases very slowly. CUCB is biased towards exploring edges which have not been triggered often. Since typical networks contain numerous edges, CUCB ends up exploring much more than necessary and results in a slow rate of decrease in regret. We observe this behaviour for other datasets as well, so we omit CUCB from the further plots. We also omit RAND from further plots to keep them simple.

\begin{figure*}[ht]
		\centering
        \subfigure[Flixster]
        {
			\includegraphics[scale=0.36]{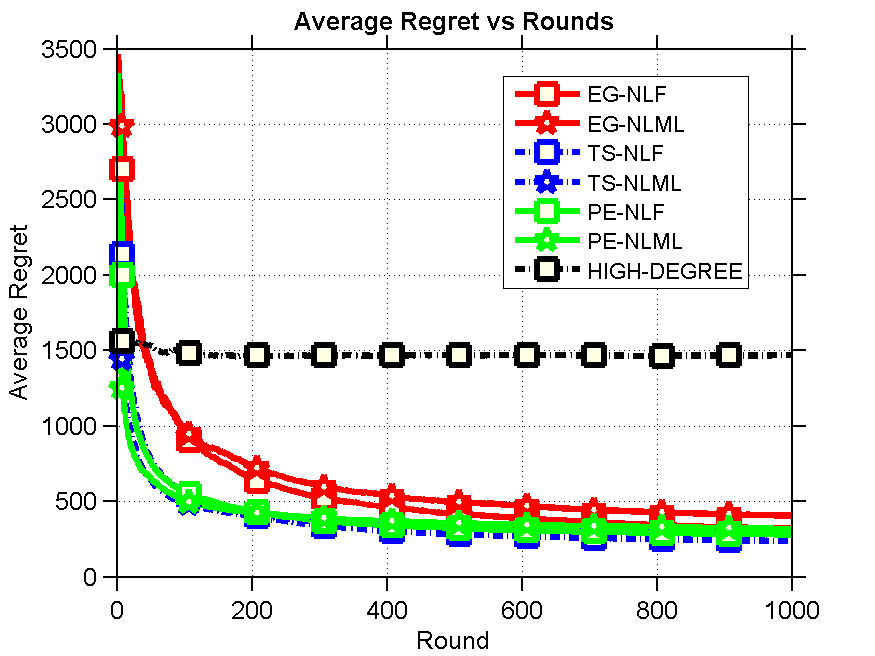}
			\label{fig:flixster-50-regret}
        }
        \subfigure[Epinions]
        {
           \includegraphics[scale=0.36]{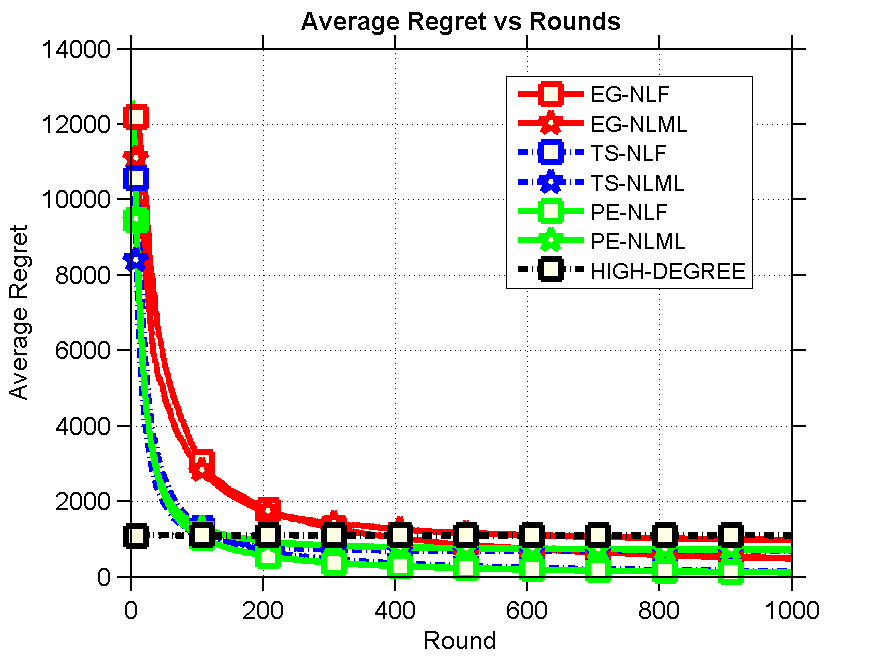}
  			\label{fig:epinions-50-regret}
        } 
        \subfigure[Flickr]
        {
           \includegraphics[scale=0.36]{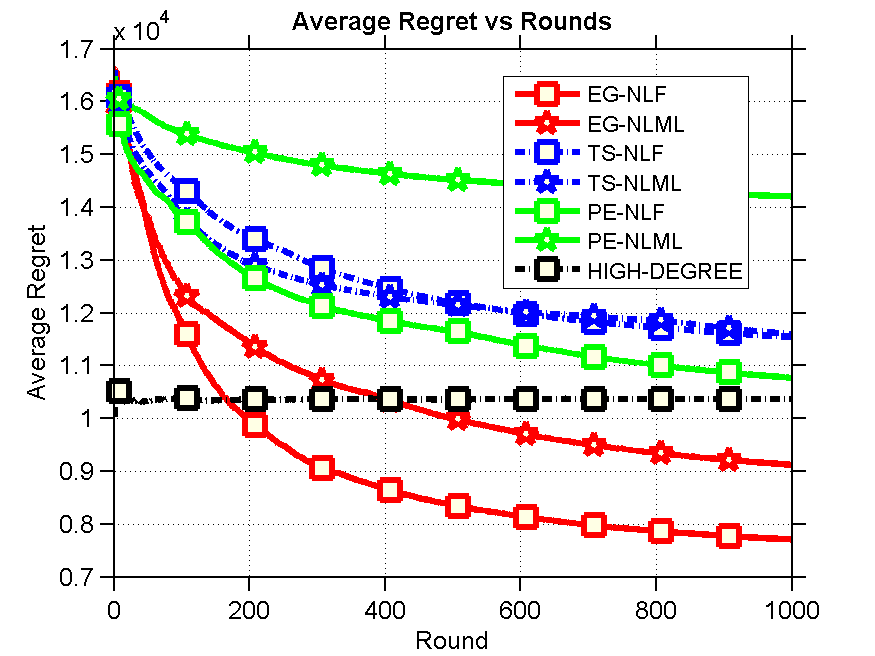}
			\label{fig:flickr-50-regret}
        } 
	    \caption{Regret vs number of rounds for different algorithms} 
            \label{fig:regret} 
\end{figure*}
To examine the effect of the feedback mechanism on regret, we plot the average regret under different feedback mechanisms in Figure~\ref{fig:nethept-50-regret}. For NetHEPT, the regret decreases quickly under both node-level feedback mechanisms and is close to that obtained with edge-level feedback. For NetHEPT with $k = 50$, the average number of active parents for a node is $1.175$. Previous work has shown that the probabilities learned from diffusion cascades are generally small~\cite{saito2011learning,goyal2010learning,netrapalli2012learning}. For example, if $p_{min} = 0$ and $p_{max}$ varies from $0.01$ to $0.2$, the failure probability $\rho$ (calculated according to the equation~\ref{eq:failure-probability}) varies from $0.0115$ to $0.2261$. This is true for all our datasets. Thus, as the number of active parents increases, credit distribution becomes more difficult and credit distribution using maximum likelihood become more effective. For all our datasets, the regret using either node-level feedback is close to that obtained using edge-level feedback mechanism. For the other datasets, to reduce clutter we just plot regret for node-level feedback mechanisms (Figure~\ref{fig:regret}). For Flixster and Epinions, both NLF and NL-ML are effective for all regret minimization algorithms with TS and PE obtaining the lowest regret. Interestingly, for  Epinions  HIGH-DEGREE is a competitive baseline and has low regret. For  Flickr, because of the large size of the graph, it is challenging to find a good seed set with partially learned probabilities. As a result, the average regret after 1000 rounds is higher than for other datasets. We observe that while both TS and PE do find a locally optimal seed set. However, because of its exploration phase, EG is able to find a much better seed set and consequently converges to a much lower regret. To verify this, we plot the relative L2 error in the edge probabilities against the number of rounds. 
\begin{figure}[!ht]
\centering
\includegraphics[scale=0.45]{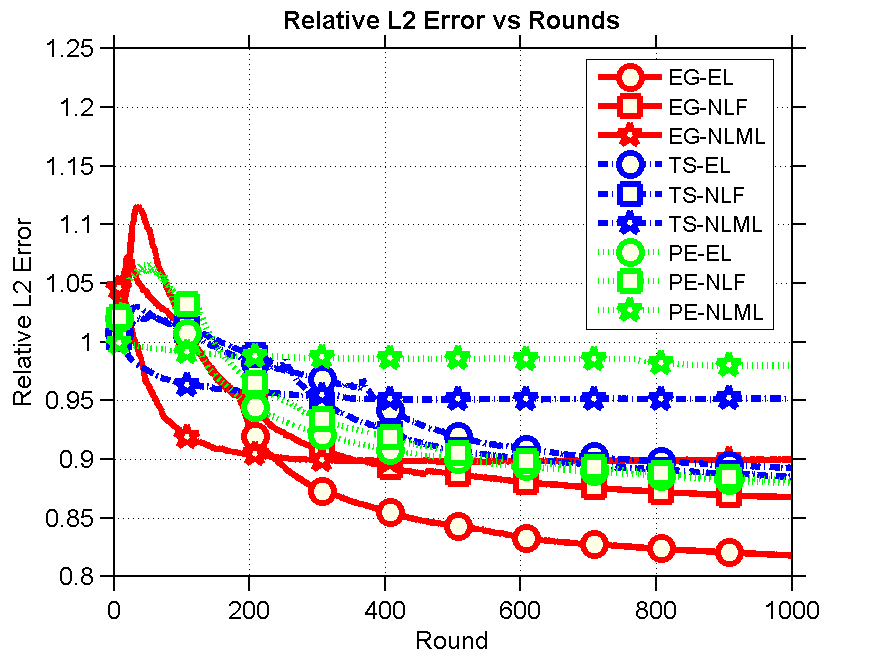}
\caption{Flickr, $k$ = 50: L2 error vs Rounds}
\label{fig:flickr-L2-error}
\end{figure}

\textbf{Quality of learning edge probabilities:} As is evident from Figure~\ref{fig:flickr-L2-error}, the mean estimates improve as the rounds progress and the relative L2 error goes down over time. This leads to better estimates of the expected spread and the quality of the chosen seeds improves. The true spread achieved thus increases and hence the average regret goes down. We see that for both PE and TS, the decrease in L2 error saturates relatively fast which implies that both of them narrow down on a seed set quickly. They subsequently stop learning about other edges in the network. In contrast, $\epsilon$-greedy does a fair bit of exploration and hence achieves a lower L2 error. 

%% file: Conclusion.tex
\vspace*{1ex}
\section{Conclusion}
\label{sec:Conclusion}
\vspace*{-4ex}
We studied the important, but under-researched problem of influence maximization when no influence probabilities or diffusion cascades are available. We adopted a combinatorial multi-armed bandit paradigm and used algorithms from the bandits literature to minimize the loss in spread due to lack of knowledge of influence probabilities. We also evaluated their empirical performance on four real datasets. It is interesting to extend the framework to learn, not just influence probabilities, but the graph structure as well.

%% file: Supplementary.tex
\appendix
\section{Proofs}
\label{sec:proofs}
\begin{theorem}
Let $p^{E}_{u_{i},v}$ and $p^{N}_{u_{i},v}$ resp. denote the probability estimates learned from edge-level and node-level feedback using maximum likelihood. We have: 
Let $p^{E}_{u_{i},v}$ and $p^{N}_{u_{i},v}$ resp. denote the probability estimates learned from edge-level and node-level feedback using maximum likelihood. We have: 
\begin{flalign*}
|p^{N}_{u_{i},v} - p^{E}_{u_{i},v}| \leq \max \bigg( & 1 - p^{E}_{u_{i},v} - \frac{1}{(p^{E}_{u_{i},v} + \failure)(P_{max})}, p^{E}_{u_{i},v} - 1 + \frac{1}{(p^{E}_{u_{i},v} + \failure)(P_{min})} \bigg) 
\end{flalign*}
where $\failure$ is the fraction of cascades  in which edge $(u_{i},v)$ is dead, over those where $v$ is active, and $P_{max}$ and $P_{min}$ are the upper bound and lower bounds on the quantity $\prod_{j \in B^{c} j \ne i} (1 - p_{u_{j},v}^{N})$. 
\begin{proof}
We want to estimate the error for probability of the edge $(u_{i},v)$ while using the maximum likelihood approach for credit distribution. Let $F_{\mathcal{E}}$ be the number of instances for which the event $\mathcal{E}$ failed. For example, $F_{u_{i},v}$ is number of times edge ($u_{i},v$) is dead and $F_{v}$ is the number of times node $v$ is inactive. Similarly, let $S_{\mathcal{E}}$ be the number of successful events and $T_{\mathcal{E}}$ be the total number of events. Clearly, $S_{\mathcal{E}} + F_{\mathcal{E}} = T_{\mathcal{E}}$. 

Let $v$ be a node under consideration at time $t$. Let $B^{c} = \{ u_{1},..u_{K^{c}} \}$ be its set of active parents (which became active at timestamp $t-1$ in the diffusion process) for cascade $c$. For our case, $S_{v} + F_{v} = S_{u_{i},v} + F_{u_{i},v} = T_{u_{i},v}$. Here, $T_{u_{i},v}$ is the number of rounds in which the edge $(u_{i},v)$ is triggered. Let $p_{u_{i},v}^{E}$ and $p_{u_{i},v}^{N}$ denote the learnt probability estimates under the edge level and node level feedback respectively. The update using edge level feedback implies $p_{u_{i},v}^{E} = \frac{S_{u_{i},v}}{T_{u_{i},v}}$. 

If $v$ isn't active at $t$, it implies that activation attempts from all active parents failed and the corresponding edge $(u_{i},v)$ is dead. If $v$ is activated, edge $(u_{i},v)$ may or may not be live. In the case of node-level feedback, we cannot observe its status. Let $S^{s}_{v}$ be the number of times node $v$ is active because of a successful activation attempt through edge $(u_{i},v)$ and let $S^{f}_{v}$ be the number of times node $v$ became active because of an active parent other than $u_{i}$ i.e. edge $(u_{i},v)$ is dead. 
We then have the following relations:
\begin{flalign}
S_{v} &= S^{s}_{v} + S^{f}_{v}  \label{preq:NL-EL-relation1} \\
S^{s}_{v} &= S_{u_{i},v}  \label{preq:NL-EL-sucess-relation} \\
F_{u_{i},v} &= S^{f}_{v} + F_{v}  \label{preq:NL-EL-relation2} 
\end{flalign}

The gradient of $\ln(LL_{v})$ wrt $p_{u_{i},v}^{N}$ can be written as:
\begin{equation}
\frac{\partial (\ln{LL_{v}})}{\partial p_{u_{i},v}^{N} } = \sum_{c = 1}^{F_{v}} \frac{-1}{1 - p_{u_{i},v}^{N}} + \sum_{c = 1}^{S_{v}}\frac{P^{c}}{1 - P^{c}(1 - p_{u_{i},v}^{N})} 
\label{preq:gradient}
\end{equation}

Here, $P^{c} = \prod_{j \in B^{c} j \ne i} (1 - p_{u_{j},v}^{N})$ i.e. it is a product over the active parents of $v$ in cascade $c$ i.e. it is the probability that in cascade $c$, all active parents (other than  $u_i$) of $v$ other failed. 

To obtain the probability estimates under node-level feedback, we set $\frac{\partial (\ln{LL_{v}})}{\partial p_{u_{i},v}^{N} } = 0$ which implies:
\begin{equation}
\frac{F_{v}}{1 - p_{u_{i},v}^{N}} = \sum_{c = 1}^{S_{v}}\frac{P^{c}}{1 - P^{c}(1 - p_{u_{i},v}^{N})}
\label{preq:gradient-solve}
\end{equation}

Let the maximum and minimum values $P^{c}$ be $P_{max}$ and $P_{min}$ respectively. Then $\frac{P^{c}(1 - p_{u_{i},v}^{N})}{1 - P^{c}(1 - p_{u_{i},v}^{N})}$ is bounded by $P^{'}_{max}$ and $P^{'}_{min}$ where $P^{'}_{max} = \frac{P_{max}(1 - p_{u_{i},v}^{N})}{1 - P_{max}(1 - p_{u_{i},v}^{N})}$ and similarly for $P^{'}_{min}$. Hence, 

\begin{equation}
\frac{S_{v}}{P^{'}_{max}} \leq \sum_{c = 1}^{S_{v}}\frac{P^{c}(1 - p_{u_{i},v}^{N})}{1 - P^{c}(1 - p_{u_{i},v}^{N})} \leq \frac{S_{v}}{P^{'}_{max}} 
\label{preq:gradient-solve-bounds}
\end{equation}

From Eq. (\ref{preq:gradient-solve}) and~\ref{preq:gradient-solve-bounds}, 
\begin{equation}
P^{'}_{min} \leq \frac{F_{v}}{S_{v}} \leq P^{'}_{max}
\label{preq:gradient-solve-nice}
\end{equation}

From Eq. (\ref{preq:NL-EL-relation1}),~\ref{preq:gradient-solve-nice} and~\ref{preq:NL-EL-relation2}
\begin{equation}
P^{'}_{min} \leq \frac{F_{u_{i},v} - S^{f}_{v}}{S_{u_{i},v} + S^{f}_{v}} \leq P^{'}_{max}
\label{preq:NL-EL-relation3}
\end{equation}

\begin{equation}
\frac{1}{P^{'}_{min} + 1} \geq p^{E}_{u_{i},v} + \frac{S^{f}_{v}}{S_{v}} \geq \frac{1}{P^{'}_{max} + 1}
\label{preq:NL-EL-relation4}
\end{equation}

Let $\failure = \frac{S^{f}_{v}}{S_{v}}$. $\failure$ depends on the structure of the network and the true probabilities. Note that $\mathbb{E}[\failure] = (1 - p^{*}_{u_{i},v})[ 1 - \prod_{j \ne i}(1 - p_{u_{j},v}^{*})]$ where $p^{*}$ denotes true probabilities. 

If $p^{N}_{u_{i},v} \geq p^{E}_{u_{i},v}$ let $p^{N}_{u_{i},v} - p^{E}_{u_{i},v} = \epsilon_{1}$. Plugging this into equation~\ref{preq:NL-EL-relation4}, we have
\begin{equation}
\epsilon_{1} \leq 1 - p^{E}_{u_{i},v} - \frac{1}{(p^{E}_{u_{i},v} + \failure)(P_{max})}
\label{preq:error-bound}
\end{equation}

If $p^{N}_{u_{i},v} \leq p^{E}_{u_{i},v}$ let $p^{E}_{u_{i},v} - p^{N}_{u_{i},v} = \epsilon_{2}$. Plugging this into equation~\ref{preq:NL-EL-relation4}, we have
\begin{equation}
\epsilon_{2} \leq p^{E}_{u_{i},v} -1 + \frac{1}{(p^{E}_{u_{i},v} + \failure)(P_{min})}
\label{preq:error-bound2}
\end{equation}

If $\epsilon$ is the error in estimation i.e. $\vert p^{N}_{u_{i},v} - p^{E}_{u_{i},v} \vert$ $\leq$ $\epsilon$, then from Eq. (\ref{preq:error-bound}) and~\ref{preq:error-bound2}, we have $\epsilon \leq max(p^{E}_{u_{i},v} -1 + \frac{1}{(p^{E}_{u_{i},v} + \failure)(P_{min})}, 1 - p^{E}_{u_{i},v} - \frac{1}{(p^{E}_{u_{i},v} + \failure)(P_{max})})$ which was to be shown. 

%In the absence of clashes i.e. where $v$ has only 1 active parent, $K^{c} = 1$ and $S2_{v} = 0$. In this case, it is easy to verify that $p_{(u_{i},v)}^{N} = p_{(u_{i},v)}^{E}$. To simplify expression, let's replace each $K^{c}$ by an average estimate $K$ i.e. $\forall c, K^{c} = K$. Equation~\ref{eq:NL-EL-relation2} then simplifies to
%\begin{equation}
%\frac{F_{(u_{i},v)} - S2_{v}}{S_{v}} = [{1 - p_{(u_{i},v)}^{N}}]\frac{K}{1 - K(1 - p_{(u_{i},v)}^{N})}
%\label{eq:NL-EL-relation3}
%\end{equation}
%
%Using equation~\ref{eq:NL-EL-sucess-relation}, 
%\begin{equation}
%\frac{1 - p_{(u_{i},v)}^{E}}{p_{(u_{i},v)}^{E}} - \frac{S2_{v}}{S_{v}} = {1 - p_{(u_{i},v)}^{N}}\frac{K}{1 - K(1 - p_{(u_{i},v)}^{N})}
%\label{eq:NL-EL-relation4}
%\end{equation}
%Here, $\frac{S2_{v}}{S_{v}}$ is a constant and depends on the true probabilities ($p^{*}$) in the network. It essentially gives the number of successful activations of $v$ not because of $u_{i}$. Specifically, $\mathbb{E}\big[ \frac{S2_{v}}{S_{v}} \big ] = (1 - p^{*}_{(u_{i},v)})[ 1 - \prod_{j \ne i}(1 - p_{(u_{j},v)}^{*})]$. 
\end{proof}
\end{theorem}
%-------------------------------------------------------------------------------------------------------- %
\begin{theorem}
If we use Eq. (7) for updating $\vec{\theta}$ with $\eta_{s}$ decreasing as $\frac{1}{\sqrt{s}}$, the following  holds:
\begin{equation}
\sum_{s=1}^{T} (L^{s}_{v}(\vectheta_{batch}) - L^{s}_{v}(\vectheta_{s}) \leq \frac{d_v \theta_{max}^{2} \sqrt{T}}{2} + (\sqrt{T} - \frac{1}{2})\gradient^{2}. 
\label{preq:OCO-specific}
\end{equation}
where $\gradient = max_{s\in[T]} ||\nabla(-L^{s}(\vec{\theta}_{s}))||$ is the maximum L2-norm of the gradient of the negative likelihood function over all rounds. 
\begin{proof}
The proof is an adaptation of the following loss result established in \cite{zinkevich2003online}:
\begin{equation}
{\it Loss\/}(T) \leq \frac{dia(F)^{2} \sqrt{T}}{2} + (\sqrt{T} - \frac{1}{2})||\nabla(c_{max})||^{2}
\label{preq:OCO- general}
\end{equation}
where $\vert \vert \nabla c_{max} \vert \vert$ is the maximum gradient obtained across the $T$ rounds in the framework of \cite{zinkevich2003online}. Turning to our setting, let the true influence probabilities lie in the range $(0,p_{max}]$, for some $p_{max}$. Then the $\theta$ values for various edges lie in the range $(0,\theta_{max})$ where $\theta_{max} = - \ln(1 - p_{max})$. Our optimization variables are $\vec{\theta}$ and the cost function $c^{s}$ in our setting is $-L^{s}_{v}$, $1 \le s \le T$. Furthermore, in our case, $dia(F) = \sqrt{d_v} \theta_{max}$ since this is the maximum distance between any two ``$\theta$-vectors'' and ${\it Loss\/}(T) = \sum_{s = 1}^{T} (L^{s}_{v}(\vec{\theta}_{batch}) - L^{s}_{v}(\vec{\theta}_{s})$. Substituting these values in Eq.~\ref{preq:OCO- general}, we obtain Eq.~\ref{preq:OCO-specific}, proving the theorem. 
\end{proof}
\end{theorem}
%--------------------------------------------------------------------------------------------------------%

%DUMP
%--------------------------------------------------------------------------------------------------------%
%\begin{reptheorem}{th:oco}
%If we use Eq.(\ref{eq:OCO-update}) for updating $\vecp$ with $\eta_{s}$ decreasing as $\frac{1}{Hs}$, the following relation holds for all $v$
%\begin{equation}
%\sum_{s=1}^{T} (L^{s}_{v}(\vecp_{batch}) - L^{s}_{v}(\vecp_{s}) \leq \frac{G^{2}}{2H}[1 + \log{T}]
%\label{eq:OCO-specific}
%\end{equation}
%\begin{proof}
%The proof is an adaptation of the loss result from \cite{hazan2007logarithmic}, reproduced below: 
%\begin{equation}
%{\it Loss\/}(T) \leq \leq \frac{G^{2}}{2H}[1 + \log{T}]
%\label{eq:OCO- general}
%\end{equation}
%where $G$ is the upper bound on the l2 norm of the gradient across rounds and $H$ is the lower bound on  hessian. Our optimization variables are $\vecp$ and the function $f^{s}$ in our setting is $-L^{s}_{v}$ where $1 \le s \le T$ and ${\it Loss\/}(T) = \sum_{s = 1}^{T} (L^{s}_{v}(\vecp_{batch}) - L^{s}_{v}(\vecp_{s})$. Substituting these values in Eq.~\ref{eq:OCO- general}, we obtain Eq.~\ref{eq:OCO-specific}, proving the theorem. 
%\end{proof}
%\end{reptheorem}
%--------------------------------------------------------------------------------------------------------%
\begin{theorem}
Let $p_{min}$ and $p_{max}$ be the minimum and maximum true influence probabilities in the network. Consider a particular cascade $c$ and any active node $v$ with $K_{c}$ active parents. The failure probability $\rho$ for under frequentist node-level feedback for node $v$ is characterized by:  
\begin{equation}
\rho \leq \frac{1}{K_{c}}(1 - p_{min})\bigg(1 - \prod_{k = 1, k \neq i}^{K_{c}}[1 - p_{max}]\bigg) + \bigg(1 - \frac{1}{K_{c}}\bigg)p_{max}. 
\label{preq:failure-probability}
\end{equation}
Suppose $\hat{\mu}_{i}^{E}$ and $\hat{\mu}_{i}^{N}$ are the inferred influence probabilities for the edge corresponding to arm $i$ using edge-level and node-level feedback respectively. Then the relative error in the learned influence probability is given by: 
\begin{equation}
\bigg\vert\frac{ \hat{\mu}_{i}^{N} - \hat{\mu}_{i}^{E}}{\hat{\mu}_{i}^{E}}\bigg\vert = \rho \bigg\vert (\frac{1}{\hat{\mu}_{i}^{E}} - 2) \bigg\vert 
\label{preq:error-in-mean}
\end{equation}
\label{th:failure-prob}
\begin{proof} 
Consider any active node $v$ with $K_{c}$ active parents. Consider updating the influence probability of the edge $(u_i, v)$. We may infer the edge $(u_i,v)$ to be live or dead. Our credit assignment makes an error when the edge is live and inferred to be dead and vice versa. Recall that all probabilities are conditioned on the fact that the node $v$ is active at time $t$ and $K_{c}$ of its parents ($u_1, ..., u_i, ..., u_{K_{c}}$) became active at time $t-1$. Let $\mathcal{E}_{d}$ ($\mathcal{E}_{l}$) be the event that the edge $(u_{i},v)$ is dead (resp., live) in the true world. Hence we can characterize the failure probability as follows: 
\begin{flalign*}
\rho &= \pr[(u_{i},v) \mbox{ inferred live}]\pr[\mathcal{E}_{d} \mid v \mbox{ is active at t}] & \\
& + \pr[(u_{i},v) \mbox{ inferred dead}]\pr[\mathcal{E}_{l} \mid  v \mbox{ is active at t}] &
\end{flalign*}
If $(u_{i},v)$ is live in the true world, then node $v$ will be active at time $t$ irrespective of the status of the edges $(u_j, v), j\in[K_c], j\ne i$. Hence, $\pr[\mathcal{E}_{l} \mid v \mbox{ is active at t}] = \pr[\mathcal{E}_{l}]$. 

By definition of independent cascade model, the statuses of edges are independent of each other. Hence, 
\begin{flalign*}
& \pr[\mathcal{E}_{d} \mid v \mbox{ is active }\mbox{at t}] = \pr[\mathcal{E}_{d} \;\wedge\; \exists \mbox{$j \ne i$ s.t.} (u_j, v) \mbox{ is live}] & \\
& \rho = \pr[(u_{i},v) \mbox{ inferred live}]\pr[\mathcal{E}_{d} \;\wedge\; \exists \mbox{$j \ne i$ s.t.} (u_j, v) \mbox{ is live}] & \\ 
& \hspace*{2ex} + \pr[(u_{i},v) \mbox{ inferred dead}]\times \pr[\mathcal{E}_{l}] &
\end{flalign*}
Let $p_{u_{j},v}$ be the true influence probability for the edge $(u_{j},v)$, $j\in[K_c]$. Thus, $\pr[\mathcal{E}_{l}] = p_{u_{i},v}$
\begin{flalign*}
& \pr[\mathcal{E}_{d} \,\wedge\, \exists \mbox{$j \ne i$ s.t.} (u_j, v) \mbox{ is live}] & \\
& = (1 - \pr[\mathcal{E}_{l}])\big[ 1 - \prod_{j = 1, j \neq i}^{K_c}[ 1 - p_{u_{j},v} ] \big] & \\
\end{flalign*}
Since one of the active nodes is chosen at random and assigned credit, $\pr[\mbox{choosing } u_{i} \mbox{ for credit}] = \pr[(u_{i},v) \mbox{ inferred live}] = \frac{1}{K_c}$. We thus obtain: 
\begin{flalign}
& \rho = \frac{1}{K_c} (1 - p_{u_{i},v}) [ 1 - \prod_{k = 1, k \neq i}^{K_c}\big[ 1 - p_{u_{k},v}] \big] +  (1 - \frac{1}{K_c}) p_{u_{i},v} & \label{preq:failure-probability-1}
\end{flalign}
Let $p_{min}$ ($p_{max}$) denote the minimum (resp. maximum) true influence probability of any edge in the network. Plugging these into Eq. (\ref{preq:failure-probability-1}) gives us the upper bound in Eq. (\ref{preq:failure-probability}), the first part of the theorem. 
Let $\hat{\mu}_{i}^{N}$ and $\hat{\mu}_{i}^{E}$ denote the mean estimates using node-level and edge-level feedback respectively. That is, they are the  influence probabilities of edge $(u_i,v)$ learned under node and edge-level feedback. We next quantify the error in $\hat{\mu}_{i}^{N}$ relative to $\hat{\mu}_{i}^{E}$. 
Let $X_{i,s}^{N}$ be the status of the edge corresponding to arm $i$ inferred using our credit assignment scheme, at round $s$. Recall that under both edge-level and node-level feedback, the mean is estimated using the frequentist approach. That is, \textbf{$\hat{\mu}_{i}^{N}$} = $\sum_{s = 1}^{T} \frac{X_{i,s}^{N}}{T_{i}}$ (similarly for edge-level feedback). Note that $X_{i,s}$ denotes the true reward (for edge level feedback) whereas $X^N_{i,s}$ denotes the inferred reward under node-level feedback, using the credit assignment scheme described earlier. Thus, for each successful true activation of arm $i$ (i.e., $X_{i,s} = 1$) we obtain $X_{i,s}^{N} = 1$ with probability $1 - \rho$ and for each unsuccessful true activation, we obtain $X_{i,s}^{N} = 1$ with probability $\rho$. Let $S_{i}$ denote the number of rounds in which the true reward $X_{i,s} = 1$. Hence, we have: 
\begin{flalign}
\hat{\mu}^E_i &= \frac{S_i}{T_i} \label{preq:edge-level-mean-estimate} \\ 
\hat{\mu}_{i}^{N} &= \frac{S_{i} (1 - \rho) + (T_{i} - S_{i})(\rho)}{T_{i}} \label{preq:node-level-mean-estimate}
\end{flalign}
The second part of the theorem, Eq.(\ref{preq:error-in-mean}), follows from Eq.(\ref{preq:edge-level-mean-estimate}) and (\ref{preq:node-level-mean-estimate}) using simple algebra.  
\end{proof}
\end{theorem}
%--------------------------------------------------------------------------------------------------------- 

%% file: Supplementary2.tex
%\appendix
\section{Effect of prior}
\label{sec:prior-effect}
\begin{figure}[!h]
\centering
\includegraphics[scale=0.45]{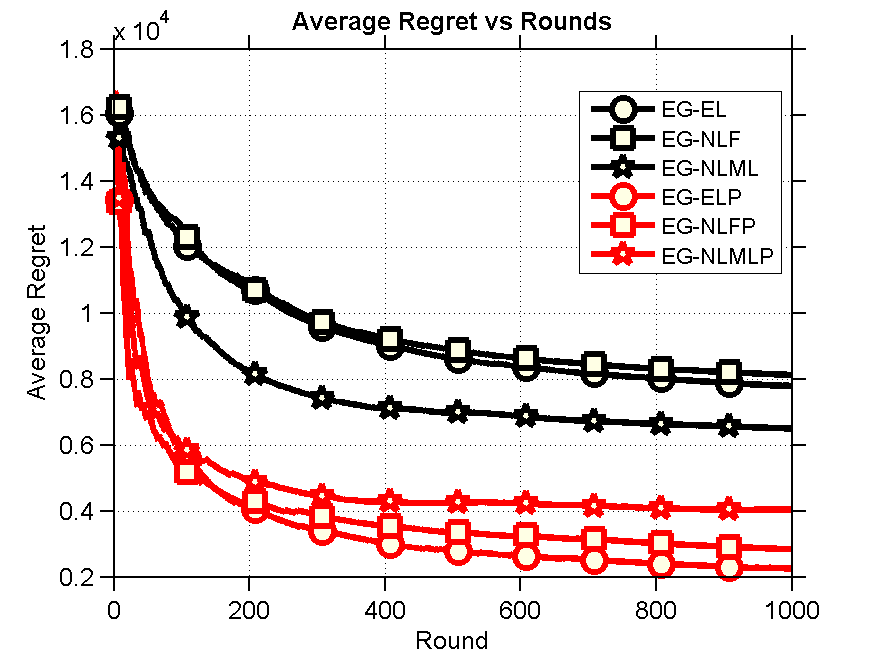}
\label{fig:flickr-50-with-prior-regret}
\caption{Effect of prior for Flickr, $k$ = 50}
\end{figure}

\begin{figure*}[ht]
\centering
        \subfigure[L2 Error]
        {
			\includegraphics[scale=0.45]{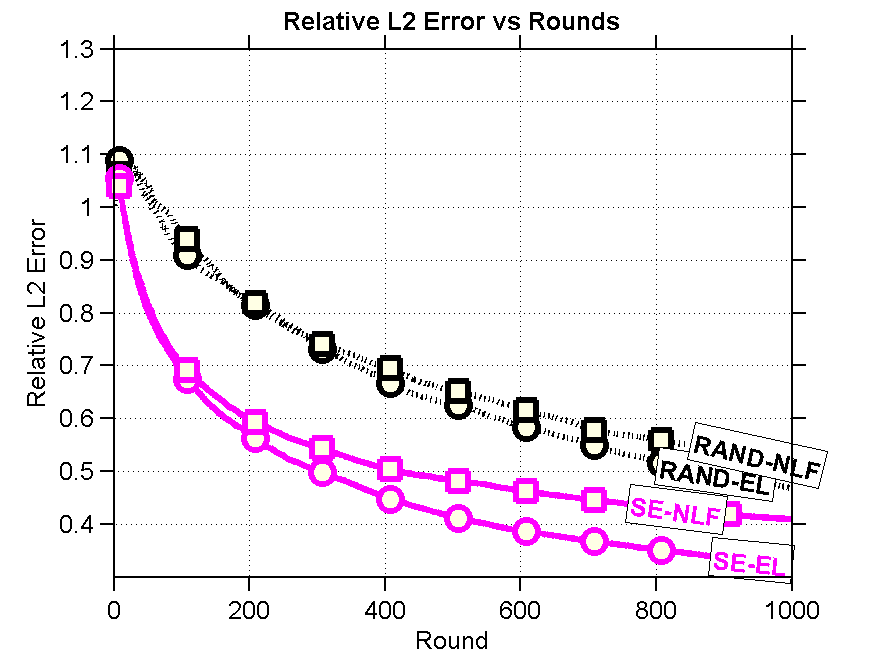}
			\label{fig:flixster-exploration-error}
        } 
        \subfigure[Fraction of edges within $10\%$ Rel Err]
        {
			\includegraphics[scale=0.45]{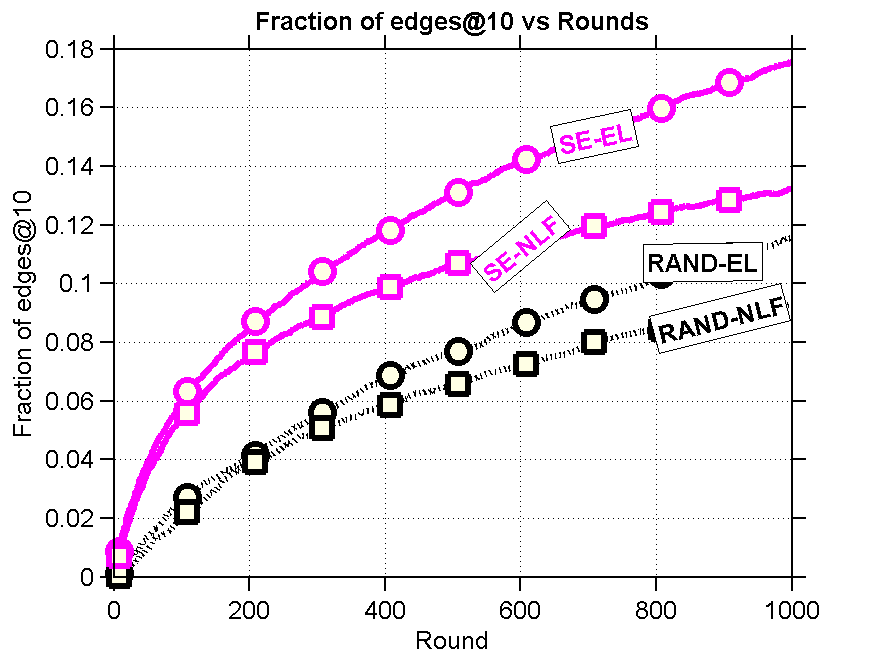}
			\label{fig:flixster-exploration-fraction}
        }
       \caption{Network Exploration for Flixster, $k$ = 50} 
   	    \label{fig:network-exploration}
\end{figure*}

For typical social networks, we may have an idea on the range of influence probabilities. E.g., we may know that the influence probabilities lie in the range of $[0.0005, 0.2]$ for a given network. If available, we can use this domain specific information to better initialize the influence probability estimates. For the maximum likelihood approach, these initial estimates can prove to be important for faster convergence of the gradient descent method. For the frequentist (both edge-level and node-level approaches), where the updates are binary i.e. the $X_{i,t}$ follow a Bernoulli distribution the initialization can be treated as a Beta prior characterized  by the parameters $\alpha$ and $\beta$ the mean of which can be given by: $\frac{\alpha}{\alpha + \beta}$.  The Bernoulli and Beta distributions are conjugate priors and the posterior follows a Beta distribution. The mean of the posterior which results in a modified update rule given by: \textbf{$\hat{\mu_{i}}$} = $\frac{\sum_{s = 1}^{t}X_{i,s} + \alpha}{T_{i,t} + \alpha + \beta}$. Hence the Beta prior parameters act like pseudo counts in the update formula. For the maximum likelihood method, we initialize the $p$ estimates randomly between $0$ and $\frac{2 \alpha}{\alpha + \beta}$. We use a prior with $\alpha = 1$ and $\beta = 19$ (similar to~\cite{lei2015online}) and show its effect  (Figure~\ref{fig:flickr-50-with-prior-regret}) on the Flickr dataset for the best performing $\epsilon$-greedy algorithm. In this figure, $ELP$ shows the regret for edge-level feedback with the prior. Similarly for the other feedback mechanisms. 

%\appendix
\section{Network Exploration}
\label{sec:nw-exp}
Instead of minimizing the regret and doing well on the IM task, one might be interested in exploring the network and obtaining good estimates of the network probabilities. We refer to this alternative task as network exploration. The objective of network exploration is to obtain good estimates of the network's influence probabilities, regardless of the loss in spread in each round and it thus requires pure exploration of the arms. Thus, we seek to minimize the error in the learned (i.e., estimated) influence probabilities $\vec{\hatmu}$ w.r.t. the true influence probabilities $\vecmu$ i.e. minimize $|| \vec{\hatmu} - \vecmu ||_{2}$. We study two exploration strategies -- \emph{random exploration}, which chooses a random superarm at each round and \emph{strategic exploration}, which chooses the superarm which leads to the triggering of a maximum number of edges which haven't been sampled sufficiently often. 

\textbf{Strategic Exploration:} Random exploration doesn't use information from previous rounds to to select seeds and explore the network. On the other hand, a pure exploitation strategy selects a seed set according to the estimated probabilities in every round. This leads to selection of a seed set which results in a high spread and consequently triggers a large set of edges. However, after some rounds, it stabilizes choosing the same/similar seed set in each round. Thus a large part of the network may remain unexplored. We combine ideas from these two extreme, and propose a strategic exploration algorithm: in each round $s$, select a seed set which will trigger the maximum number of edges that have not been explored sufficiently often until this round. We instantiate this intuition below. 

Recall $T_{i}$ is the number of times arm $i$ (edge $(u_i,v)$) has been triggered, equivalently, number of times $u_i$ was active in the $T$ cascades. Writing this in explicit notation, let $T^s_{(u,v)}$ be the number of times the edge $(u,v)$ has been triggered in the cascades $1$ through $s$, $s\in[T]$. 
Define $value(u) := \sum_{v\in N^{out}(u)} \frac{1}{T^{s}_{(u,v)} + 1}$. Higher the value of a node, the more unexplored (or less frequently explored) out-edges it has. Define \emph{value-spread} of a set $S\subset V$ exactly as the expected spread $\sigma(S)$ but instead of counting activated nodes, we add up their values. Then, we can choose seeds with the maximum marginal value-spread gain w.r.t. previously chosen seeds. It is intuitively clear that this strategy will choose seeds which will result in a large number of unexplored (or less often explored) edges to be explored in the next round. We call this strategic exploration (SE). It should be noted that the value of each node is dynamically updated by SE across rounds so it effectively should result in maximizing the amount of exploration across the network. 

We show results on the Flixster dataset. Figure~\ref{fig:flixster-exploration-error} shows the L2 error obtained by using Random Exploration and Strategic Exploration strategies, coupled with Edge level feedback and the frequentist node-level feedback mechanisms. First, we can see that strategic exploration is better than just choosing nodes at random because it incorporates feedback from the previous rounds and explicitly tries to avoid those edges which have been sampled (often). As expected, edge level feedback shows the faster decrease in error. In Figure~\ref{fig:flixster-exploration-fraction}, we plot the fraction of edges which are within a relative error of $10\%$ of their true probabilities. Since we have the flexibility to generate cascades to learn about the hitherto unexplored parts of the network, our network exploration algorithms can lead to a far lesser sample complexity as compared to algorithms which try to learn the probabilities from a given set of cascades. This is similar to the benefits obtained using active learning as compared to supervised learning.